\documentclass[fleqn,multphys,vecphys]{svmult}

\usepackage{makeidx}   
\usepackage{graphicx}  
\makeindex             
\usepackage{hyperref}  

\newcommand{\iDev}[1]{#1}
\newcommand{\iName}[1]{#1}
\newcommand{\iStreet}[1]{#1}
\newcommand{\iPostcode}[1]{#1}
\newcommand{\iCity}[1]{#1}
\newcommand{\iCountry}[1]{#1}
\newcommand{\email}[1]{\texttt{#1}}

\def\abstractname{Abstract.}%

\begin{document}

\title*{$^{59}$Co, $^{23}$Na, and $^{1}$H NMR Studies of Double-Layer Hydrated Superconductors Na$_{x}$CoO$_{2}\cdot y$H$_{2}$O}          
\toctitle{$^{59}$Co, $^{23}$Na, and $^{1}$H NMR Studies of Double-Layer Hydrated Superconductors Na$_{x}$CoO$_{2}\cdot y$H$_{2}$O}        
\titlerunning{NMR Studies of Hydrated Superconductors Na$_{x}$CoO$_{2}\cdot y$H$_{2}$O}                                

\author{Y. Itoh\inst{1}, H. Ohta\inst{1}, C. Michioka\inst{1}, M. Kato\inst{2}, and K. Yoshimura \inst{1}                     
}
\authorrunning{Y. Itoh, H. Ohta, C. Michioka, and K. Yoshimura} 

\institute{\iDev{Department of Chemistry, Graduate School of Science},                  
\iName{Kyoto University}, \newline
\iStreet{Oiwake-cho, Kitashirakawa, Sakyo-ku},
\iPostcode{606-8502}
\iCity{Kyoto},
\iCountry{Japan}\newline
\email{kyhv@kuchem.kyoto-u.ac.jp}
\and 
\iDev{Department of Molecular Science and Technology, Faculty of Engineering},
\iName{Doshisha University},\newline
\iStreet{Tatara-Miyakodani},
\iPostcode{610-0394}
\iCity{Kyotanabe},
\iCountry{Japan}\newline
\email{}
}

\maketitle

 \begin{abstract}
                 We present our NMR studies of 
double-layer hydrated cobalt oxides Na$_{x}$CoO$_{2}\cdot y$H$_{2}$O ($x\approx$0.35, $y\approx$1.3)
with various $T_\mathrm{c}$ = 0 $-$ 4.8 K and magnetic transition temperatures. 
High-resolution $^{1}$H NMR spectrum served as an evidence for the existence of H$_{3}$O$^{+}$ oxonium ions.
$^{23}$Na nuclear spin-lattice relaxation rates served to detect local field fluctuations sensitive to $T_\mathrm{c}$.   
$^{59}$Co nuclear quadrupole resonance (NQR) spectra served to classify the various $T_\mathrm{c}$ samples. 
From the classification by $^{59}$Co NQR frequency,
the double-layer hydrated compounds were found to have
two superconducting phases closely located to a magnetic phase. 
In the normal state and at a magnetic field in the $ab$-plane, 
two $^{59}$Co NMR signals with different Knight shifts   
and different $^{59}$Co nuclear spin-lattice relaxation times $^{59}T_{1}$ were observed.   
The two $^{59}$Co NMR signals suggest
magnetic disproportionation of two Co sites or inplane (XY) anisotropy of a single Co site. 
Non Korringa behavior and 
power law behavior in zero-field NQR 1/$^{59}T_{1}$ above and below $T_\mathrm{c}$ 
suggest non-Fermi liquid and unconventional superconductivity. 
 \end{abstract}

\section{Introduction}
\begin{figure}
\includegraphics[width=11cm]{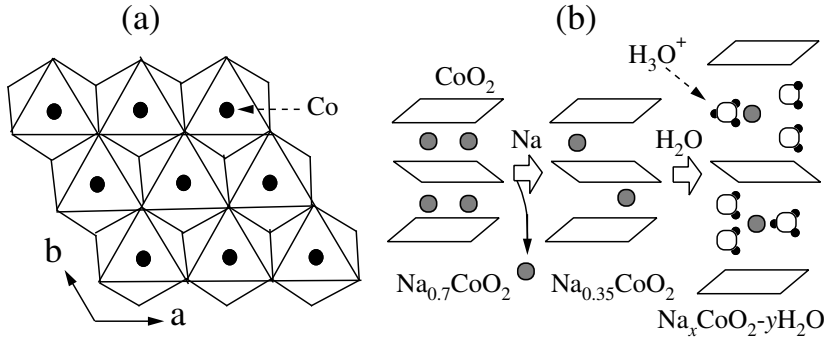}
    \caption{\label{fig:crystal}
    (a)The top view of a CoO$_{2}$ plane. (b)The side views of the crystal structures
    of a parent Na$_{0.7}$CoO$_{2}$, a Na-deintercalated Na$_{0.35}$CoO$_{2}$ 
    and a ${\it bi}$layer hydrated Na$_{0.35}$CoO$_{2}\cdot y$H$_{2}$O. 
    Open arrows indicate soft chemical treatment.}
      \label{faksim}
\end{figure}

Na$_{0.7}$CoO$_{2}$ is a high-performance thermoelectric power material \cite{Terasaki}. 
The crystal structure is constructed from the stacking of
a Co triangular lattice in the layered edge-sharing CoO$_{2}$ octahedrons
and of a Na layer.  
The discovery of superconductivity 
through soft chemical treatment of sodium de-intercalation and hydration
has renewed our interests in spin and charge frustration effects 
on the CoO$_{2}$ triangular lattice \cite{Takada}. 
Figs.~\ref{fig:crystal}(a) and~\ref{fig:crystal}(b) illustrate the crystal structure and the soft chemical treatment, respectively.
The double-layer hydrated cobalt oxides Na$_{x}$CoO$_{2}\cdot y$H$_{2}$O ($x\approx$0.35, $y\approx$1.3) are  triangular-lattice superconductors with the optimal $T_\mathrm{c}\approx$ 5 K \cite{Takada}. 
Except the charge ordered insulator Na$_{x}$CoO$_{2}$ with $x\approx$ 1/2,
Na$_{x}$CoO$_{2}$ and the hydrated compounds are itinerant electronic systems.
For $x\approx$0.7, $A-$type spin fluctuations
(in-plane ferromagnetic and out-of-plane antiferromagnetic spin fluctuations) 
\cite{Boothroyd1, Boothroyd2}, 
the motion of Na ions \cite{Gavilano1, Gavilano2} and 
Co charge disproportionation \cite{Gavilano2, Imai, Mukhamedshin_Na, Mukhamedshin_Co} 
have been reported.  

For anhydrated Na$_{x}$CoO$_{2}$, the Na concentration changes the
carrier doping level of a CoO$_{2}$ plane, which develops the electronic state \cite{FooPRL}.
For double-layer hydrated Na$_{x}$CoO$_{2}\cdot y$H$_{2}$O, the Na concentration 
is not a unique parameter of $T_\mathrm{c}$ \cite{Takada_oxonium, Sakurai, Milne}. 
Since no one has ever observed superconductivity of the anhydrated compounds,
the carrier doping level is not a unique parameter of $T_\mathrm{c}$. 
The chemical diversity of the site ordering of Na ions \cite{Cava} and 
the charge compensation by oxonium ions H$_{3}$O$^{+}$ \cite{Takada_oxonium} 
may realize a delicate electronic system on the triangular lattice. 
The effect of the intercalated water molecules  
on the electronic state of the CoO$_{2}$ plane is poorly understood. 

The followings have been debated so far, 
whether the normal-state spin fluctuations are ferromagnetic, antiferromagnetic or the other type,  
whether or not the superconductivity occurs in the vicinity of the magnetic instability, 
whether the charge fluctuations play a key role in the superconductivity,  
how different the triangular-lattice superconductivity is from the square-lattice one,  
what the pairing symmetry is.

In this paper, we present the highlights of our NMR studies and findings for the double-layer hydrated compounds; 
the various $T_\mathrm{c}$ samples synthesized in controllable way, 
$^{1}$H NMR evidence for the existence of  the H$_{3}$O$^{+}$ oxonium ions,
$T_\mathrm{c}$-sensitive $^{23}$Na nuclear spin-lattice relaxation,  
two superconducting phases classified by $^{59}$Co nuclear quadrupole resonance (NQR) frequency,
a magnetic phase located between the two superconducting phases,
in-plane two components in $^{59}$Co NMR Knight shift and nuclear spin-lattice relaxation, 
non-Korringa behavior above $T_\mathrm{c}$,
and power law behaviors in nuclear spin-lattice relaxation rates.   

\section{Synthesis and experiments} 
	Polycrystalline samples of the starting compound Na$_{0.7}$CoO$_{2}$ were synthesized by conventional solid-state reaction methods. 
The powdered samples of Na$_{0.7}$CoO$_{2}$ were immersed in Br$_{2}$/CH$_{3}$CN solution to deintercalate Na$^{+}$ ions and then in distilled water to intercalate H$_{2}$O molecules \cite{Ohta, Ohta2, OhtaLD} 
or  in Br$_{2}$/H$_{2}$O solution to make ion-exhange reaction of Na and H$_{2}$O \cite{Barnes, OhtaUP}. 
After the powders of Na$_{0.35}$CoO$_{2}\cdot y$H$_{2}$O were separated from the solution
by filtration, 
they were exposed in various humidity air. 
For the synthesized samples, 
we observed various duration (keeping time in the humidity-controlled chamber in a daily basis) effects on $T_\mathrm{c}$ \cite{Ohta, Ohta2, OhtaLD, OhtaUP, OhtaNa, Michioka_CoNQR}. 

$^{59}$Co (spin $I$ = 7/2), $^{23}$Na (spin $I$ = 3/2) and $^{1}$H (spin $I$ = 1/2) NMR experiments were performed by a pulsed NMR spectrometer for the powder samples. 
The nuclear spin-lattice relaxation times were measured by inversion recovery techniques. 
    
\section{Experimental results}
\subsection{Magnetic phase diagram classified by $^{59}$Co NQR}
\begin{figure}
\includegraphics[width=10cm]{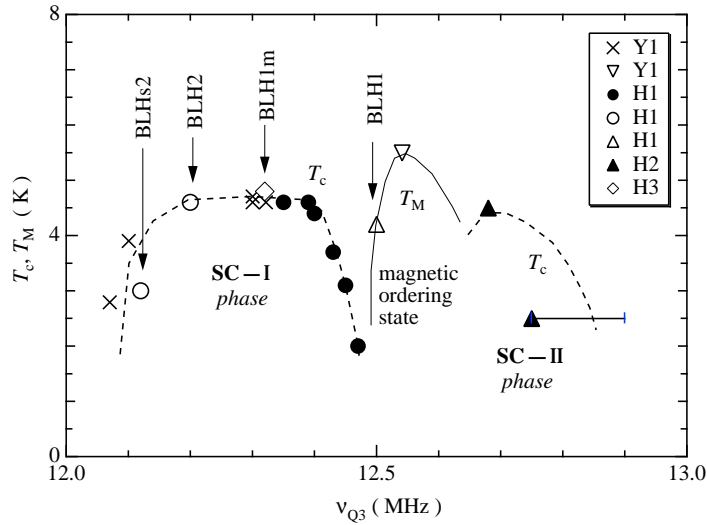}
    \caption{\label{fig:phasediagram}
    Magnetic phase diagram of $T_\mathrm{c}$ and $T_\mathrm{M}$
    plotted against $^{59}$Co NQR frequency $\nu_\mathrm{Q3}$ at 10 K. 
We reproduced the data of Y1 from \cite{Ihara_MO}, 
H1 from \cite{Ohta, Michioka_CoNQR}, 
H2 from \cite{OhtaUP},
and H3 from \cite{Michioka_CoNMR}.   
The solid and dashed curves are guides for eyes. 
 We call two superconducting phases characterized by $\nu_\mathrm{Q3} < $ 12. 5 MHz and by $\nu_\mathrm{Q3} > $ 12. 6 MHz {\it ${\bf SC-I}$ phase} and {\it ${\bf SC-II}$ phase}, respectively.
 The samples denoted by BLH ($bi$layer hydrates) are referred to the NMR experiments. }
      \label{faksim}
\end{figure}

Zero-field NQR
is sensitive to local crystal structure and charge distribution around the nuclear site. 
Although no distinct changes in X-ray diffraction patterns have been observed 
for the various double-layer hydrated compounds, 
systematic changes in the peak values of $^{59}$Co NQR spectra were observed 
\cite{Ohta, Michioka_CoNQR, Ihara_MO, IharaFull}.
Thus, we present the results of $^{59}$Co NQR spectrum measurments to characterize the samples
and to classify the electronic states. 

For an electric field gradient tensor $V_{\alpha\beta}$ 
($\alpha\beta$ = xx, yy, and zz, principal axis directions)
with an asymmetry parameter $\eta$[$\equiv$($V_{xx}$-$V_{yy}$)/$V_{zz}$] $<$ 1,
three transition lines of  $^{59}$Co (spin $I$ = 7/2) NQR should be observed as
$\nu_{Q1}$ ($I_{z}$ = $\pm$3/2 $\leftrightarrow \pm$1/2),
$\nu_{Q2}$ ($I_{z}$ = $\pm$5/2 $\leftrightarrow \pm$3/2),
and 
$\nu_{Q3}$ ($I_{z}$ = $\pm$7/2 $\leftrightarrow \pm$5/2). 
For $\eta \ll$ 1,
one easily finds $\nu_{Q3}$ $\approx$ 1.5$\nu_{Q2}$ $\approx$ 3$\nu_{Q1}$.

In Fig.~\ref{fig:phasediagram}, superconducting transition temperatures $T_\mathrm{c}$ 
and magnetic transition temperatures $T_\mathrm{M}$ are
plotted against $^{59}$Co NQR frequency $\nu_\mathrm{Q3}$ at 10 K
 \cite{Ohta, OhtaUP, Michioka_CoNQR, Ihara_MO, Michioka_CoNMR}.
The existence of two superconducting phases were confirmed. 
We call the two superconducting phases at $\nu_\mathrm{Q3} < $ 12. 5 MHz and at $\nu_\mathrm{Q3} > $ 12. 65 MHz {\it ${\bf SC-I}$ phase} and {\it ${\bf SC-II}$ phase}, respectively.
In the magnetic phase diagram of Fig.~\ref{fig:phasediagram}, 
the optimal samples of $T_\mathrm{c}$ = 4.5 $-$ 4.8 K in the ${\bf SC-I}$ $phase$ are located 
in 12.1 MHz $<$ $\nu_\mathrm{Q3}$ $\leq$ 12.4 MHz. 
The magnetic ordering phase is located between two superconducting phases. 
Thus, the magnetic instability and the superconductivity occur closely to each other. 
 
It should be emphasized that for the samples with 12.5 MHz $< \nu_\mathrm{Q3} < $ 12. 65 MHz,
a possibility of charge-density-wave (CDW) ordering is excluded by the temperature dependence
of both $^{59}$Co NQR spectra of $\nu_{Q2}$ and $\nu_{Q3}$ \cite{Michioka_CoNQR, Ihara_MO}. 
The CDW ordering at a transition temperature $T_\mathrm{M}$ must result in the broad NQR line widths of
$\Delta \nu_{Q3}$ $\approx$ 1.5$\Delta \nu_{Q2}$ $\approx$ 3$\Delta \nu_{Q1}$.  
However, $\Delta \nu_{Q2} \gg \Delta \nu_{Q3}$ was observed below $T_{M}$ \cite{Michioka_CoNQR, Ihara_MO}. 
Thus, an internal magnetic field due to magnetic ordering is concluded.  
Here, since no divergence behavior at $T_\mathrm{M}$ was observed 
in nuclear spin-lattice relaxation rates, 
this magnetic ordering is unconventional \cite{OhtaNa, Michioka_CoNQR}.   
No wipeout effect below $T_\mathrm{M}$ excludes the existence of slow modes 
due to spin glass transition \cite{Michioka_CoNQR, Ihara_MO}.   

It should be noted that the experimental magnetic phase diagram 
against $\nu_{Q3}$ in Fig.~\ref{fig:phasediagram} is similar to the theoretical one against the thickness of a CoO$_{2}$ layer \cite{Mochizuki1, Mochizuki2}. 
In the theoretical phase diagram, 
two superconducting phases are separated through a magnetic ordering phase and  
result from two competing Fermi surfaces of $a_{1g}$ cylinders and $\acute{e}_{g}$ hole pockets. 
Two superconducting phases should have different pairing symmetry \cite{Mochizuki1, Mochizuki2}. 
We present the NMR studies of the ${\bf SC-I}$ phase below.   

\subsection{$^{1}$H NMR of oxonium ion}
\begin{figure}
\includegraphics[width=8cm]{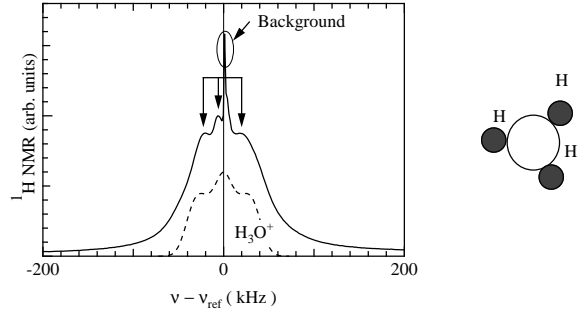}
    \caption{\label{fig:proton}
$^{1}$H proton NMR spectrum at 200 MHz for a double-layered hydrated sample at room temperature. The dashed curve is a numerical simulation of $^{1}$H NMR powder pattern for H$_{3}$O$^{+}$
illustrated in the right figure.}
      \label{faksim}
\end{figure}

In the double-layer hydrated superconductors, not only the concentration of Na$^{+}$ ions \cite{FooPRL}
but also the H$_{3}$O$^{+}$ ions \cite{Takada_oxonium, Sakurai, Milne} play key roles in
the occurrence of superconductivity.  
The existence of the oxonium ions H$_{3}$O$^{+}$ was evidenced by the observation
of a bending mode and stretching modes of H$_{3}$O$^{+}$ ions 
using Ramann spectroscopy \cite{Takada_oxonium}. 
NMR is also a powerful technique to detect such a local structure. 

Fig.~\ref{fig:proton} shows a high resolution Fourier-transformed $^{1}$H NMR spectrum at 200 MHz for a double-layered hydrated sample in the ${\bf SC-I}$ phase at room temperature. 
We observed one peak and two shoulders in the $^{1}$H NMR spectrum.
For a frozen water molecule, a Pake doublet of two peaks should be observed \cite{Abragam}. 
In Fig.~\ref{fig:proton}, a numerical simulation for an oxonium ion H$_{3}$O$^{+}$ 
with triangular coordination of three protons
is illustrated by a dashed curve \cite{Abragam}.
Similarity between the multiple structure in the $^{1}$H NMR spectrum
and the numerical simulation in Fig.~\ref{fig:proton} indicates the existence of the oxonium ion H$_{3}$O$^{+}$
in the double-layer hydrated compound. 
It is a future work to estimate how much H$_{3}$O$^{+}$ ions and H$_{2}$O molecules are involved
in each sample. 

\subsection{$A$-type spin fluctuations via $^{23}$Na NMR}
Since Na ions are located between CoO$_{2}$ layers,
it is expected that Na nuclear spins can probe interlayer correlations.
\begin{figure}
\includegraphics[width=12cm]{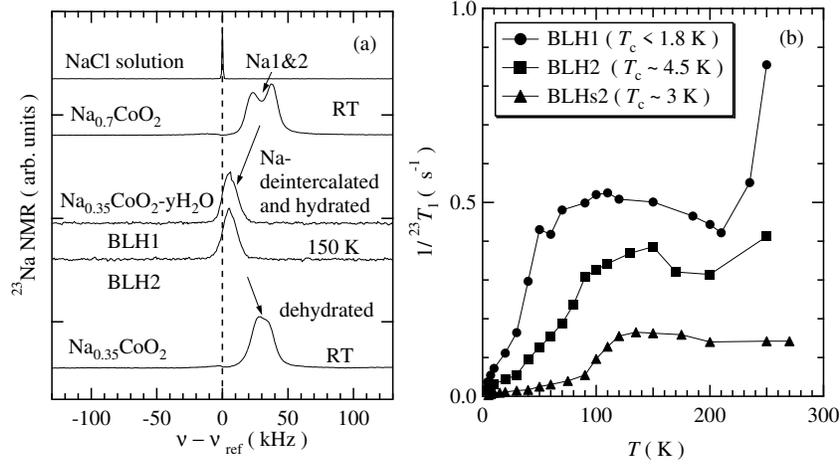}
    \caption{\label{fig:NaNMR}
    (a)Fourier-transformed $^{23}$Na NMR frequency spectra of Na$_{0.7}$CoO$_{2}$, 
double-layer hydrated Na$_{0.35}$CoO$_{2}\cdot y$H$_{2}$O 
(BLH1 and BLH2 denoted in Fig.~\ref{fig:phasediagram}), and a dehydrated Na$_{0.35}$CoO$_{2}$. 
The dashed line is $\nu_\mathrm{ref}$ = 84.2875 MHz ($\nu_\mathrm{ref}$=$^{23}\gamma_\mathrm{n}H$, $^{23}\gamma_\mathrm{n}$ = 11.262 MHz/T and $H \approx$ 7.484 T). 
(b)Temperature dependence of $^{23}$Na nuclear spin-lattice relaxation rates 1/$^{23}T_{1}$ 
for double-layer hydrated Na$_{0.35}$CoO$_{2}\cdot y$H$_{2}$O 
(BLH1, BLH2, and BLHs2 denoted in Fig.~\ref{fig:phasediagram}).}
      \label{faksim}
\end{figure}

\begin{figure}
\includegraphics[width=12cm]{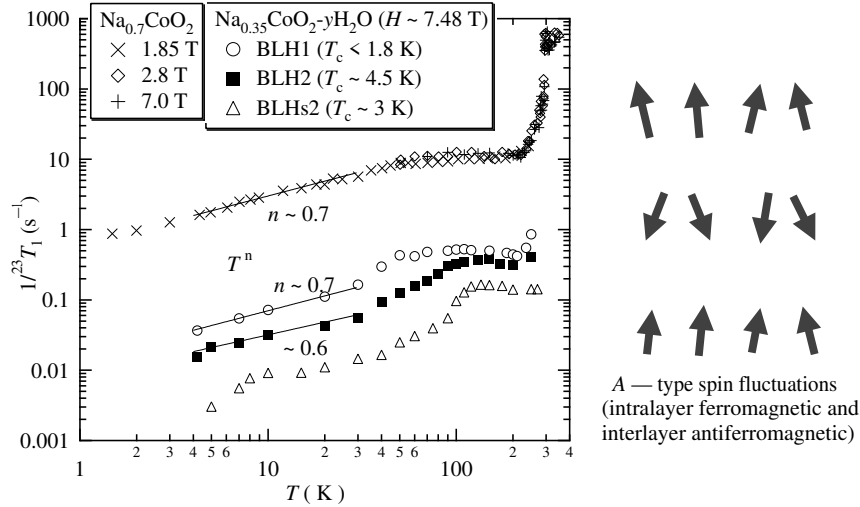}
    \caption{\label{fig:LogLogNaT1}
    Log-log plots of $^{23}$Na nuclear spin-lattice rate 1/$^{23}T_{1}$ against temperature for Na$_{0.7}$CoO$_{2}$ \cite{Gavilano1, Ihara}, double-layer hydrated Na$_{0.35}$CoO$_{2}\cdot y$H$_{2}$O (BLH1, BLH2, and BLHs2 denoted in Fig.~\ref{fig:phasediagram}). 
    The solid lines are fits  by power laws of $T^{n}$.
    The right figure illustrates a snapshot of $A$-type spin fluctuations. }
      \label{faksim}
\end{figure}
 
 Fig.~\ref{fig:NaNMR}(a) shows the central transition lines ($I_{z}$ = $\pm$1/2 $\leftrightarrow \mp$1/2) 
 of the Fourier transformed $^{23}$Na (nuclear spin $I$ = 3/2) NMR spectra 
for Na$_{0.7}$CoO$_{2}$, a double-layer hydrated non-superconducting BLH1 ($T_\mathrm{c} <$ 1.8 K) and a superconducting BLH2  ($T_\mathrm{c} \approx$ 4.5 K), both of which are denoted in Fig.~\ref{fig:phasediagram}, 
and a dehydrated Na$_{0.35}$CoO$_{2}$ \cite{Ohta2, OhtaNa}. 
The dehydrated Na$_{0.35}$CoO$_{2}$ was obtained by heating a double-layer hydrate 
at about 250 $^{\circ}$C.  
In Na$_{0.7}$CoO$_{2}$, 
the Na NMR lines are affected by large Knight shift and electric quadrupole shift.
For the double-layer hydrates BLH1 and BLH2, however, the $^{23}$Na NMR lines show small Knight shifts. 
For the Na$_{0.35}$CoO$_{2}$ dehydrated from Na$_{0.35}$CoO$_{2}\cdot y$H$_{2}$O,
the $^{23}$Na NMR lines show large Knight shift once again. 
The recovery of Knight shift by dehydration indicates that not the Na deficiency but the intercalated water molecule blocks the Co-to-Na hyperfine field. 

	Fig.~\ref{fig:NaNMR}(b) shows the temperature dependence of the $^{23}$Na nuclear spin-lattice relaxation rate 1/$^{23}T_{1}$ for BLH1, BLH2 and BLHs2 ($T_\mathrm{c} \approx$ 3 K) \cite{Ohta2, OhtaNa, Yoshimura}. 
1/$^{23}T_{1}$ largely decreases with duration from BLH1 to BLH2 and then BLHs2. 
The local field fluctuations probed by 1/$^{23}T_{1}$ \cite{Moriya1, Moriya2}
are sensitive to the duration effect.  
	 
	Fig.~\ref{fig:LogLogNaT1} shows the log-log plots of 1/$^{23}T_{1}$ 
for the parent Na$_{0.7}$CoO$_{2}$ 
which are reproduced from \cite{Gavilano1,Ihara} 
and for the double-layer hydrated BLH1, BLH2 and BLHs2 \cite{Ohta2, OhtaNa}. 
At low temperatures in the normal states above $T_\mathrm{c}$, 
1/$^{23}T_{1}$'s show the power-law behaviors (solid lines) of $T^{n}$ with $n \approx$ 0.7 and 0.6 for BLH1 and BLH2, respectively. 
These non-Korringa behaviors indicate that the samples are not conventional Fermi liquid systems. 
The magnitude of 1/$^{23}T_{1}$ significantly decreases from the non-hydrated to the double-layer hydrated samples. 
	
	The temperature dependence of 1/$^{23}T_{1}$ in BLH1 is nearly the same as that in the parent Na$_{0.7}$CoO$_{2}$ \cite{OhtaNa}. 
Both systems show the power law of $T^{0.7}$ in 1/$^{23}T_{1}$ at $T <$ 20 K and the rapid increase at $T >$ 200 K. 
The electron spin dynamics probed at the Na site in the double-layer hydrates is similar to that in Na$_{0.7}$CoO$_{2}$. 
The $A$-type spin fluctuations, i.e. intra-plane ferromagnetic and inter-plane antiferromagnetic fluctuations, are observed in Na$_{0.7}$CoO$_{2}$
by inelastic neutron scattering experiments \cite{Boothroyd1, Boothroyd2}. 
The intercalated water molecules may block the inter-plane antiferromagnetic couplings. Nearly ferromagnetic intra-plane spin fluctuations may persist in the double-layer hydrates.

\section{$^{59}$Co NMR for ${\bf SC-I}$ phase} 
\subsection{Normal state}
\begin{figure}
\includegraphics[width=12cm]{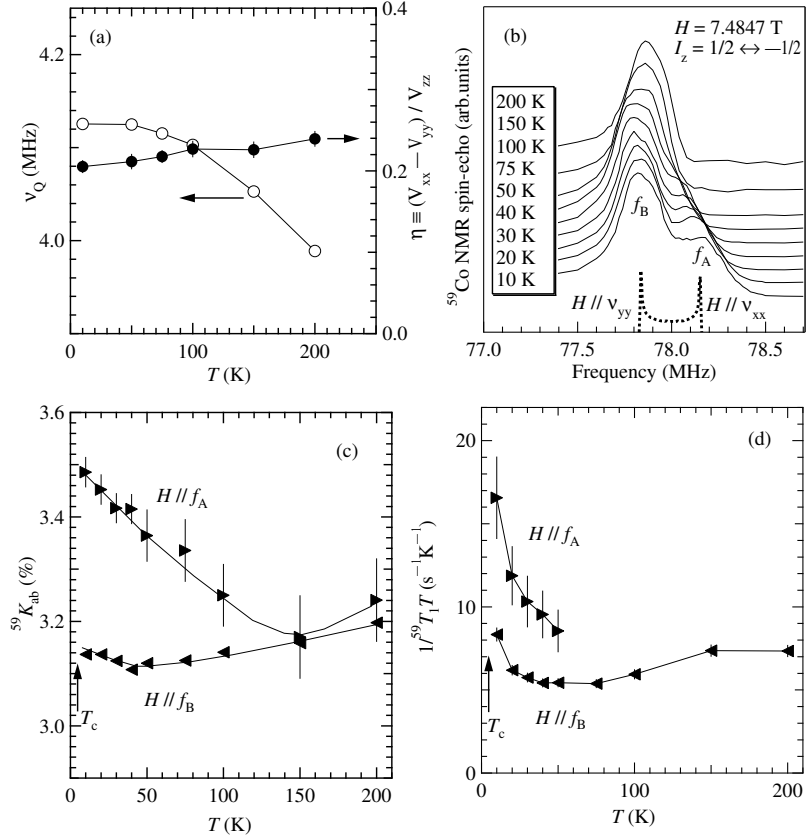}
    \caption{\label{fig:CoNMR}
    $^{59}$Co NQR and NMR for an optimal sample (BLH1m in Fig.~\ref{fig:phasediagram}) of $T_\mathrm{c}$ = 4.8 K.
    (a)$^{59}$Co nuclear quadrupole resonance frequency $\nu_{Q}$ and the asymmetry parameter $\eta$. 
    (b)The central transition lines ($I_{z}$ = 1/2 $\leftrightarrow -$1/2) of $^{59}$Co NMR spectra at $H // ab$ plane
and a numerical simulation (dotted curve) of two dimensional powder pattern of the central transition
including anisotropic Knight shift of $K_{x} > K_{y}$.
    (c)$^{59}$Co Knight shifts $K_\mathrm{A}$ and $K_\mathrm{B}$ at $f_\mathrm{A}$ and $f_\mathrm{B}$.  
    (d)$^{59}$Co nuclear spin-lattice relaxation rates divided by temperature 1/$T_{1}T$ at $f_\mathrm{A}$ and
     $f_\mathrm{B}$.}
      \label{faksim}
\end{figure}

From $^{59}$Co NQR and NMR experiments
for an optimal sample of $T_\mathrm{c}$ = 4.8 K (BLH1m denoted in Fig.~\ref{fig:phasediagram}), 
we obtained the significant results that 
the electric field gradient tensor $V_{\alpha \beta}$ at the $^{59}$Co nuclear site is asymmetric ($\eta \approx$ 0.2),
that  the in-plane $^{59}$Co Knight shift has two components 
(inplane-anisotropic $K_{x} \neq K_{y}$ or magnetic disproportionation),
and that the $^{59}$Co nuclear spin-lattice relaxation rate 1/$^{59}T_{1}$ 
is different at the two signals \cite{Michioka_CoNMR}. 

Fig.~\ref{fig:CoNMR}(a) shows the temperature dependence of $^{59}$Co nuclear quadrupole resonance frequency $\nu_{Q}$ and the asymmetry parameter $\eta$ \cite{Michioka_CoNMR}. 
The temperature dependence of $\nu_{Q}$ in Fig.~\ref{fig:CoNMR}(a) is widely observed in many materials. 
The finite asymmetry parameter $\eta \approx$ 0.2 suggests an unique underlying electronic wave funtion
and/or charge distribution at the Co site. 

Fig.~\ref{fig:CoNMR}(b) shows the temperature dependence of the central transition  
 ($I_{z}$ = $\pm$1/2 $\leftrightarrow \mp$1/2) of $^{59}$Co NMR spectra 
 at $H$(=7.4847 T) $// ab$-plane \cite{Michioka_CoNMR}.
For the $^{59}$Co NMR experiments, we prepared 
the powder sample Na$_{0.35}$CoO$_{2}\cdot y$H$_{2}$O 
oriented by a magnetic field of $H \approx$ 7.5 T 
in Fluorinert FC70 (melting point of 248 K). 
Stycast 1266 was the worst for the hydrated samples. 
Hexane was much better than Stycast 1266 and actually applied to NMR experiments \cite{Kato}
but hard to handle.
Fluorinert was the best. 
The $ab$-planes of powder grains are aligned to the external magnetic field.  
Thus, the observed NMR spectrum should be a two dimensional powder pattern. 

The dotted curve in Fig.~\ref{fig:CoNMR}(b) is a numerical simulation of two dimensional powder pattern
including anisotropic Knight shift of $K_{x} > K_{y}$. 
Here we define the Knight shift  $K_{x}$ for the line $H // \nu_{xx}$
and $K_{y}$ for the line $H // \nu_{yy}$ of 
the central transition ($I_{z}$ = $\pm$1/2 $\leftrightarrow \mp$1/2).
The full-swept $^{59}$Co NMR spectra and their field dependences were well reproduced 
from two dimensional powder pattern due to quadrupole shifts with second order perturbation 
and anisotropic Knight shifts \cite{Kato}. 
In principle, the asymmetric quadrupole shift with $\eta \approx$ 0.2 yields two peaks in the central transition
but could not reproduce the actual width 
between the two resonance peaks ($f_\mathrm{A}$ and $f_\mathrm{B}$) in Fig.~\ref{fig:CoNMR}(b). 
From the magnetic field dependence of the two peaks, 
two signals have nearly the same quadruple shifts \cite{Kato}. 
Since twice magnetic field yields twice split width between the two resonance peaks in Fig.~\ref{fig:CoNMR}(b),  
we should introduce at least two components of Knight shift \cite{Kato}.  
Since imperfect orientation of powder grains easily suppresses the singularity at $H // \nu_{xx}$,
then the disagreement of the intensity ratio of $f_\mathrm{A}$ to $f_\mathrm{B}$
in the NMR spectrum  
may not be significant. 
However, if the intensity difference of $f_\mathrm{A}$ and $f_\mathrm{B}$ is intrinsic, 
one must introduce two magnetic Co sites with nearly the same quadruple shifts. 
The charge disproportionation was observed 
in the parent Na$_{0.7}$CoO$_{2}$ \cite{Gavilano2, Imai, Mukhamedshin_Na, Mukhamedshin_Co}.
Then, magnetic disproportionation may occur in the double-layer hydrates
in that case. 

After the second order correction of quadrupole shifts based on $\nu_{Q}$ and $\eta$ in Fig.~\ref{fig:CoNMR}(a),
we estimated the respective Knight shifts $K_\mathrm{A}$ and $K_\mathrm{B}$ 
at the lines $f_\mathrm{A}$ and $f_\mathrm{B}$ in Fig.~\ref{fig:CoNMR}(b)
\cite{Michioka_CoNMR}.
Fig.~\ref{fig:CoNMR}(c) shows the temperature dependences of the Knight shifts $K_\mathrm{A}$ and $K_\mathrm{B}$ \cite{Michioka_CoNMR}. 
The Knight shift $K_\mathrm{A}$ shows a Curie-Weiss-type upturn below 50 K, 
being similar to a low temperature bulk magnetic susceptibility \cite{Kato}.
Thus,  a Curie-Weiss behavior in the magnetic susceptibility below 50 K is intrinsic.  

Fig.~\ref{fig:CoNMR}(d) shows the temperature dependence of 1/$^{59}T_{1}T$ 
at $f_\mathrm{A}$ and $f_\mathrm{B}$ \cite{Michioka_CoNMR}. 
The difference in 1/$^{59}T_{1}T$  indicates
inplane (XY)-anisotropy or disproportionation in the Co electron spin fluctuations. 

\subsection{Superconducting state}
\begin{figure}
\includegraphics[width=13cm]{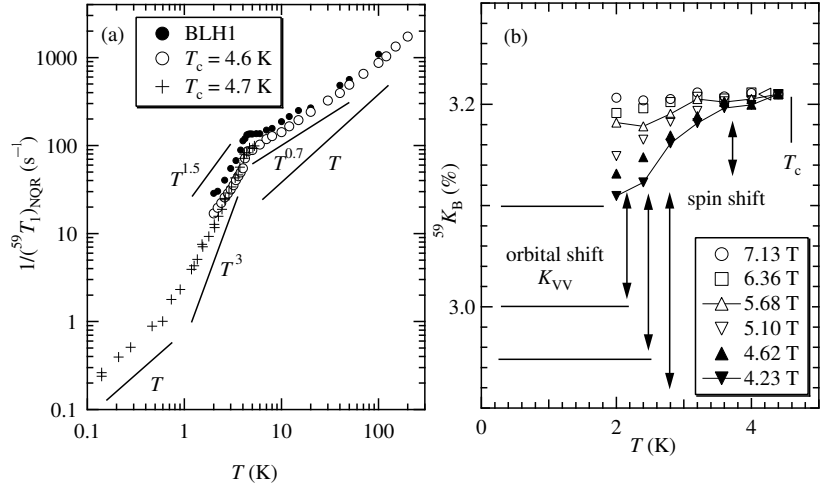}
    \caption{\label{fig:CoSCT1}
    (a)Zero-field $^{59}$Co nuclear spin-lattice relaxation rates 1/($^{59}T_{1}$)$_\mathrm{NQR}$  
    for BLH1 and BLH sample of $T_\mathrm{c}$ = 4.6 K \cite{Michioka_CoNQR} and the sample of $T_\mathrm{c}$ = 4.7 K  \cite{Ishida}.
    (b)Magnetic field dependece of the superconducting $^{59}$Co Knight shift $K_\mathrm{B}$ for the optimal sample of $T_\mathrm{c}$ = 4.7 K reproduced from \cite{Kato}. 
    The horizontal lines indicate the orbital shifts $K_\mathrm{VV}$ = 2.9, 2.95, 3.0, and 3.1 $\%$ reported so far. The arrows indicate the respective spin shifts $K_\mathrm{spin}$. }
      \label{faksim}
\end{figure}

Fig.~\ref{fig:CoSCT1}(a) shows zero-field $^{59}$Co nuclear spin-lattice relaxation rates 1/($^{59}T_{1}$)$_\mathrm{NQR}$  
 for BLH1,  the sample of $T_\mathrm{c}$ = 4.6 K \cite{Michioka_CoNQR} 
 and the sample of $T_\mathrm{c}$ = 4.7 K \cite{Ishida}.
No Korringa behavior in 1/($^{59}T_{1}$)$_\mathrm{NQR}$ just above $T_\mathrm{c}$ and no coherence peak  just below $T_\mathrm{c}$ 
suggest non-Fermi liquid and unconventional superconductivity. 
 The power law behavior of 1/($^{59}T_{1}$)$_\mathrm{NQR}$  below $T_\mathrm{c}$ indicates
 the existence of line nodes on a superconducting gap parameter. 
In a triangular lattice, such a line-nodal order parameter is expected for $d_{xy}$-wave, $p$-wave
and $f$-wave pairing \cite{Dagotto, Yanase1, Yanase2, Yanase3}
and extended $s$-wave inter-band pairing \cite{Kuroki}. 

Fig.~\ref{fig:CoSCT1}(b) shows the magnetic field dependence of the superconducting $^{59}$Co Knight shift $K_\mathrm{B}$ for the optimal sample with $T_\mathrm{c}$ = 4.7 K \cite{Kato}. 
    The horizontal bars indicate the reported orbital shifts 
$K_\mathrm{VV}$ = 2.7 (out of the figure scale) \cite{Kobayashi}, 2.9 \cite{Mukhamedshin_Co}, 2.95 \cite{Zheng}, 3.0 \cite{Michioka_CoNMR}, and 3.1 $\%$ \cite{Kato}. 
The Knight shift $K$ is expressed by the sum of the spin shift $K_\mathrm{spin}$ and the Van Vleck orbital shift $K_\mathrm{VV}$.
Then, the arrows indicate the respective spin shifts $K_\mathrm{spin}$. 
If  $K_\mathrm{VV}$ = 2.9 $\%$, the finite $K_\mathrm{spin}$ indicates a spin-triplet pairing.
If  $K_\mathrm{VV}$ = 3.1 $\%$, the diminished $K_\mathrm{spin}$ indicates a spin-singlet pairing.
Thus, the choice of $K_\mathrm{VV}$ changes the conclusion for the issue whether
the spin singlet or triplet is realized. 
  
The existing data on the impurity Ga- and Ir-substituion effects \cite{Yokoi} support non-$s$-wave pairing.
Although any impurity substitution effects on $T_\mathrm{c}$ as Anderson localization effect \cite{Guo} could not serve as the tests of  identification of any pairing symmetry,
weak impurity potential scattering for the line-nodal pairing could account for the observed weak suppression of  $T_\mathrm{c}$.    

\section{Conclusion} 
Double-layer hydrated cobalt oxides Na$_{x}$CoO$_{2}\cdot y$H$_{2}$O were found to have two superconducting phases labeled by $^{59}$Co NQR frequency $\nu_\mathrm{Q3}$. 
Non-superconducting magnetic phase is located between two phases
in the $\nu_\mathrm{Q3}$ classification. 
The peculiar electronic states, 
magnetic disproportionation or inplane (XY) anisotropy of $^{59}$Co local spin susceptibility, 
was observed for the ${\bf SC-I}$ $phase$.  
Although the superconducting gap parameter with line nodes seems to be established, 
there does not seem to be a robust answer to a question which is realized,  a spin singlet or triplet state.
 
\section*{Acknowledgement}
We thank  
Y. Yanase, M. Mochizuki, M. Ogata, J. L. Gavilano, J. Haase, T. Imai, and M. Takigawa for their fruitful discussions.
This study was supported by a Grant-in-Aid 
for Science Research on Priority Area,
"Invention of anomalous quantum materials" from the Ministry of 
Education, Science, Sports and Culture of Japan (Grant No. 16076210).

\end{document}